# Intelligent Implementation Processor Design for Oracle Distributed Databases System

Hassen Fadoua, Grissa Touzi Amel

Université Tunis El Manar , LIPAH, FST, Tunisia

Université Tunis El Manar ,ENIT, LIPAH,FST, Tunisia
{hassen.fadoua@gmail.com;amel.touzi@enit.rnu.tn}

**Abstract** . Despite the increasing need for modeling and implementing Distributed Databases (DDB), distributed database management systems are still quite far from helping the designer to directly implement its BDD. Indeed, the fundamental principle of implementation of a DDB is to make the database appear as a centralized database, providing series of transparencies, something that is not provided directly by the current DDBMS. We focus in this work on Oracle DBMS which, despite its market dominance, offers only a few logical mechanisms to implement distribution. To remedy this problem, we propose a new architecture of DDBMS Oracle. The idea is based on extending it by an intelligent layer that provides: 1) creation of different types of fragmentation through a GUI for defining different sites geographically dispersed, 2) allocation and replication of DB. The system must automatically generate SQL scripts for each site of the original configuration.

## 1 Introduction

The organizational evolution of companies and institutions that rely on computer systems to manage their data has always been hampered by centralized structures already installed, this architecture does not satisfy the need for autonomy and evolution of the organization because it requires a permanent return to the central entity, which leads to a huge waste of time and overwhelmed at work. Need to have a management system of databases that can handle these "data islands" in harmony with all the functional of starting system is one of the motivations that underlie the development of distributed databases.

Unfortunately, existing DDBMS have several limitations. The world's leading providers of DBMS such as Oracle, offer only few partitioning commands in console mode. The tool for Oracle DDB implementation, known as Oracle Partitioning [7], remains limited toward the expectations of end users and DBA. This tool has several limitations: 1) it does not automatically ensure the transparency of data distribution, and 2) it does not directly represent the basic concepts of distributed database (DDB) as fragmentation, etc.  The designer must describe the DDB necessary scripts to manually distribute its data to produce results consistent with its distributed design 3) It cannot handle cross sites referential integrity constraints 4) It does not cover the calculations and updates requiring access to multiple sites at once.

Consequently, the design and implementation of a DDB has never been an easy task especially when dealing with huge models and while trying to satisfy strict user expectations. In this context, we can mention the work of Rim [8] and Hassen [3] who proposed an expert system to help designing the DDB. These tools are more concerned to suggest distribution of data on different sites, regardless of the heavy task left to the designer to implement this DDB on different sites or the validation process of fragmentation if the user decides to change its design in response to new needs.

In this paper, we propose a new approach to assist design and implementation of DDB. This approach was validated through the design and implementation of an assistance tool that provides a graphical interface for different types of fragmentation, allocation and duplication along with validation at each step of the process. Then, the system automatically generates SQL scripts for each site of the initial configuration. We prove that the proposed tool can be implemented as an intelligent layer to any existing DDBMS.

Besides this introduction, this paper includes five sections. Section 2 presents the basic concepts of DDBS. Section 3 presents an example of implementation of BDD, illustrating the problems and limitations of existing DDBMS. Section 4 presents our motivation for this work. Section 5 presents our new approach for Oracle DDBS. Section 5 presents the validation of our approach by providing the platform. Finally, we conclude by evaluating this work and proposing some future perspectives of it.

## 2    Summary of Oracle Distributed Database System

Oracle announced distributed DBMS capabilities in 1987, but largely as marketing ploy. The first Oracle product to reasonably support distributed database processing is Oracle 7, which has been in the market since 1993.[1]
A distributed database appears to a user as a single database but is, in fact, a set of databases stored on multiple computers. The data on several computers can be simultaneously accessed and modified using a network. Each database server in the distributed database is controlled by its local DBMS, and each cooperates to maintain the consistency of the global database. [4]
Oracle supports two types of distributed databases. In a system based on homogeneous distributed data, all databases are Oracle database. In a system of distributed heterogeneous database, at least one of the databases is not an Oracle database.[6]

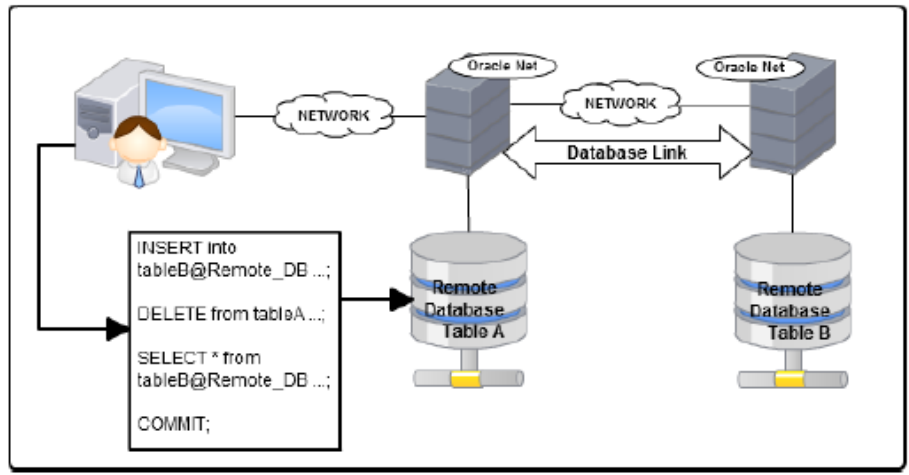

**Fig. 1**. Sample of remote and local database calls using Oracle DB links.

### 2.2 Allocation mechanism in Oracle

Like any commercial DDBMS, Oracle does not accept the type of fragmentation mechanism, although the administrator can manually allocate DB data to produce similar results. This has the effect of shifting the responsibility under the auspices of the end user, who must know that a table has been fragmented and convert this knowledge into the application. In other words, the Oracle DDBMS does not ensure transparency of the distribution, while it allows location transparency [8].

In order to ensure transparency, designer must stick to following steps:

1. User accounts creation among sites.
2. Bi-directional links creation between different sites (using oracle CREATE DBLINK command).
3. Local schema implementation on each site
4. Synonyms definition to ensure location transparency
5. View and/or materialized views creation and/or snapshots to insure fragmentation independent schema. On each materialized view and snapshot definition, we have to specify update mode (asynchronous, synchronous) and refresh delay in accordance with application need
6. Stored procedure definition, as PL/SQL script, on each update operation in a way to make data fragmentation and duplication automated and transparent.
7. As a DBMS can ensure only local data integrity, designer must define PL/SQL triggers that allow checking distributed data integrity among DDB.

## 3   Sample DDB implementation under Oracle

In this section, we implement an example of a database under RDBMS Oracle 11g as a sample. [4]

**Example**

Three institutions of the University of Tunis El Manar: National Engineering School of Tunis (NSET), Faculty of Mathematical, Physical and Natural Sciences of Tunis (FST) and Faculty of Economic Sciences and Management of Tunis (FESMT) have decided to pool their libraries and service loans, to enable all students to borrow books in all the libraries of the participating institutions. Joint management of libraries and borrowing is done by a database distributed over 3 sites (Site1 = NSET, Site2 = FST and Site3 = FESMT), the global schema is as follows:

```
EMPLOYEE (SSN,emp_ fname, emp_ lname, Address, Status, Assignment)
STUDENT (NCE, stud_fname, stu_lname, Address, Institution, Class, nb_borrow)
BOOKS (Id_book, Title, editor, Year, Area, Stock, website)
AUTHORS (Id_book, au_lname, au_fname)
Loans (Id_book, NCE, date_borrows, return_date)
```

The management of this application is based on the following assumptions:

1. An employee is assigned to a single site.
2. A student is enrolled in a single institution, but can borrow from all libraries.
3. A book borrowed from a library is returned to the same library.
4. The nb_borrows field of STUDENT relation is used to limit the number of books borrowed by a student simultaneously on all libraries. It is updated at each loan and each return, regardless of the lending library.
5. Each institution manages its own students.
6. Each library manages its staff and works it holds.

To illustrate the necessary very long task for implement our DDB, we present in the following a part of the necessary script for implement the table STUDENT.

```
--------------------------------
-- DDL for DB Link BB1
--------------------------------------------------------------------------
 CREATE DATABASE LINK "ENIT_dblink"
  CONNECT TO "ROOT" IDENTIFIED BY VALUES 'root'
  USING '(DESCRIPTION= (ADDRESS=  (PROTOCOL=TCP)  (HOST=127.0.0.1) (PORT=1521))
(CONNECT_DATA=  (SERVICE_NAME=ENIT)))';
---------------------------------------------------
-- DDL for Table STUDENT_FST
---------------------------------------------------
  CREATE TABLE  STUDENT_FST  ( NCE  NUMBER, ST_FNAME  VARCHAR2(200),
ST_FLNAME  VARCHAR2(200), ADRESS VARCHAR2(200 CHAR),
CLASS  NUMBER, CURSUS  VARCHAR2(200 CHAR),Constraint PK11 primary key (NCE)  ) ;
--------------------------------------------------------------------------
```

```
-- DDL for Synonym ENIT
------------------------------------------------------------------------
CREATE SYNONYM STUDENT_ENIT FOR STUDENT_ENIT@ENIT;
------------------------------------------------------------------------
-- DDL for materialized view STUDENT
------------------------------------------------------------------------
create materialized view STUDENT
refresh complete
start with sysdate
 next sysdate + 7
 as
(SELECT * from FOR STUDENT_ENIT@ENIT) UNION
(SELECT * from FOR STUDENT_ENIT@FST) UNION
(SELECT * from FOR STUDENT_ENIT@FSEGT)
------------------------------------------------------------------------
-- DDL for trigger insetSudent
------------------------------------------------------------------------
CREATE OR REPLACE trigger insertStudent
before insert on Student
for each rowdeclare
excep exception ;
nbTuples number ;
begin
nbTuples :=0;
select count (*) into nbTuples from Etudiant where
NCE=:new. NCE ;
i f ( nbTuples != 1) then
raise excep ;
else
IF : new .Institution := "ENIT" THEN
INSERT INTO Etudiant_ENIT (NCE,ST_FNAME, ST_LNAME, ADRESS, Institution,"CLASS",
CURSUS) VALUE
( : new . NCE , : new. ST_FNAME,: new. ST_LNAME, : new . adress ,
 new . Institution,: new . "CLASS" ,: new . cursus )
INSERT INTO Student_lib( NCE , Nb_borrow,ST_FNAME, ST_LNAME )
VALUE( : new . NCE , : new. Nb_borrow, : new. ST_FNAME,: new. ST_LNAME )
ELSIF : new .Institution :="FST" THEN
INSERT INTO Etudiant_FST
(NCE,ST_FNAME, ST_LNAME, ADRESS, Institution
    "CLASS", CURSUS) VALUE
( : new . NCE , : new. ST_FNAME,: new. ST_LNAME, : new . adress ,
 new . Institution,: new . "CLASS" ,: new . cursus )
INSERT INTO Student_lib
( NCE , Nb_borrow,ST_FNAME, ST_LNAME )
VALUE( : new . NCE , : new. Nb_borrow, : new. ST_FNAME,: new. ST_LNAME )
```

```
ELSIF : new .Institution :="FSEGT" THEN
INSERT INTO Etudiant_FSEGT
(NCE,ST_FNAME, ST_LNAME, ADRESS, Institution
    "CLASS", CURSUS) VALUE
( : new . NCE , : new. ST_FNAME,: new. ST_LNAME, : new . adress ,
 new . Institution,: new . "CLASS" ,: new . cursus )
INSERT INTO Student_lib
( NCE , Nb_borrow,ST_FNAME, ST_LNAME )
VALUE( : new . NCE , : new. Nb_borrow, : new. ST_FNAME,: new. ST_LNAME )
ELSE
RETURN 'The university name is invalid'
END IF
END IF
when excep then
raise_application_error
(–20009 , 'constraint violation' )
END;
```

## 4    Motivation

As described previously, DDB are still facing the following issues:

1. DDB design is not an easy task. Multiple criteria must be considered on this sensitive operation: Sites number, user needs and frequent queries. Designer must establish a compromise between data duplication and performances cost of update and select queries. He must find out relationship to fragment, to replicate and update type to consider on each synchronous or asynchronous relationship.
2. DDB implementation is still a heavy task nowadays especially with huge databases called from numerous sites. Actually this implementation, executed manually by designer, have to make a DDB look like a centralize BD, by ensuring a transparency list. This is not yet affordable automatically on current DDBMS.

In the following, we propose, a new architecture of DDBMS Oracle. The idea is based on extending it by an intelligent layer that provides: 1) creation of different types of fragmentation through a GUI for defining different sites geographically dispersed, 2) allocation and replication of DB. The system must automatically generate SQL scripts for each site of the original configuration.

## 5    New Approach proposals for Oracle DDBS

### 5.1    Objectives of our approach

Ideally, the new layer must satisfy the following objectives:

1. Design help for distributed schema: The layer must provide the designer with a friendly and productive interface that allows him to represent the draft of the design into a comprehensive and accessible format for review and collaboration. Fields, tables, sites suggestion lists and work tools (fragmentation and replication) must be provided to designer to ease schema graphical description and avoid additional task complication.
2. Automated implementation of design schema: Once distribution schema is established and validated by the designer along with the wizard assistance, the component "Script generator" must afford the ability to translate accurately described distribution policy to valid SQL scripts. Generated scripts can be directly executed on sites from the layer if access are already prepared, or give deliverable files to transmit to each site administrator.
3. Ensure the integrity constraints and inter-site calculations: The fragmentation horizontal and vertical fragmentation can corrupt relations starting point for the majority of DBMS (including Oracle) do not handle the constraints of integrities between two distinct sites. It is proposed to substitute the integrity check of the local DBMS triggers that maintain the uniqueness of the primary key between sites, and the validity of the foreign keys between two or more remote fragments
4. Ensure transparency toward the end-user: Creating views that combine data from remote fragments automatically. This reconstruction is an essential constituent of the transparency of the distribution.

### 5.2    Suggested layer architecture:

The architecture of our Intelligent DDB is illustrated in Figure 2 depicts the architecture of implemented layer. Assistance in the process of distribution is made by all of the following:

1. Access to centralized database to distribute
2. DB link creation
3. Horizontal, vertical and nested fragmentation
4. Fragmentation result validation
5. Data replication

At the end of the process, two options are afforded to execute scripts, depending on afforded preconditions: 1) Automatically: If design environment has valid access to remote sites, the layer executes scripts on each remote site. 2) Manually: User transfer files using an external tool and takes in charge then execution on remote sites.

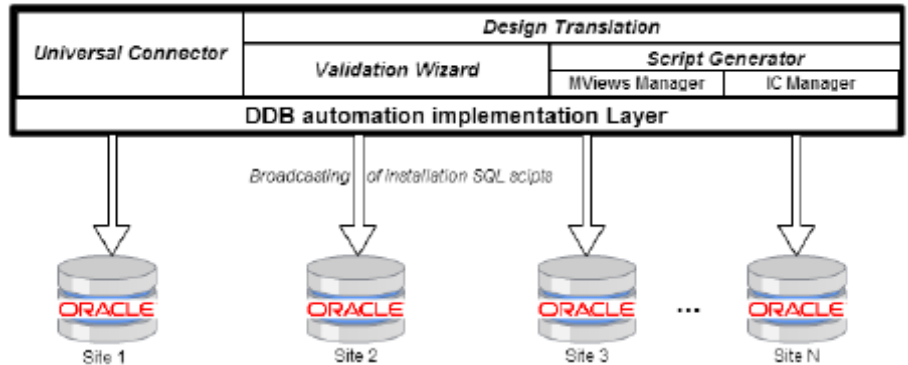

**Fig. 2.** Proposed layer architecture.

The operating principle is as follows:

Begin
1. Select database fragment
2. Enter the list of distribution sites
3. Retrieve the list of relationships based on initial schema
4. While "Complete Fragmentation" is false
    a. Select the table fragment
    b. Choose the type of fragmentation (A)
        i. If Horizontal Fragmentation
            1. Select column fragmentation
            2. Affect the value of fragmentation for each site
        ii. ELSE IF Vertical Fragmentation
            1. Name the fragment
            2. Select the columns of the fragment
            3. Select the host site of the vertical fragment
        iii. If Hybrid Fragmentation
            1. Treat the hybrid fragment
    c. Validate the resulting fragmentation
    d. Show validation report
    e. If validation is negative,return in (4.b)
    f. If the resulting fragment has foreign keys
        i.perform the derivative fragmentation
4. END WHILE
5. For each site
    a. Generate scripts for creating database links from other sites
    b. Generate Scripts fragments of this site (tables and CRUD Procs)
    d. Generate materialized views
    e. Write the script in a file name of the site
END

## 6    Intelligent-DDB

Distribution wizard "Intelligent-DDB" is intended to help users graphically distribute a centralized DB, supports the creation of DB links, horizontal, vertical, hybrid and derived fragmentation and replication. The final result is a set of SQL scripts to run on each site.

To implement our tool, we used Microsoft Windows Seven software environment. Simulation nodes in network, were made by installing two virtual machines (Oracle Virtual Box) on the chosen host. The development environment is apprehended DotNet framework 4.5 (CSharp).

Intelligent-DDB provides designers with multiple screens. We illustrate some examples as follows:

After welcome screen and tool introduction and interactive help access, user accesses the connection panel to identify target centralized database (Fig.3).

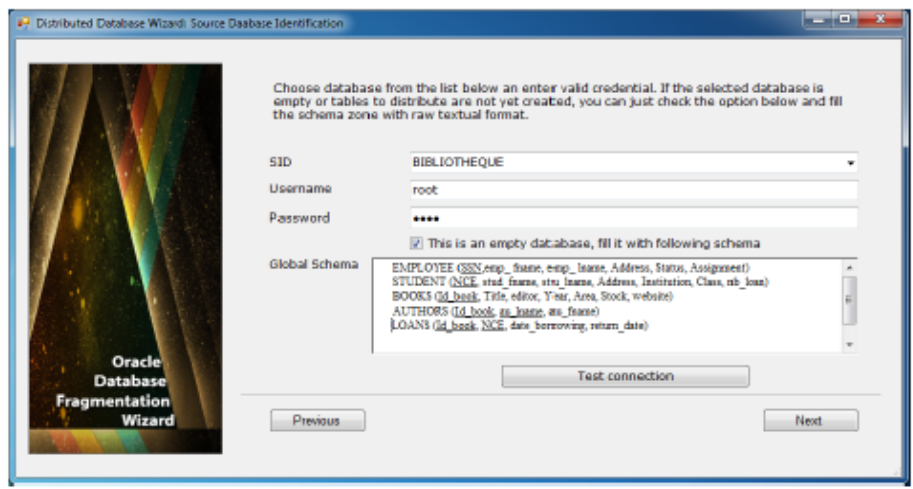

Fig. 3. Connection panel

On successful connection test, next screen is just a popup asking for the number of sites on the distribution. Then, a visual map is displayed with raw nodes. Designer must identify each site with network address (either a name or an ip), a logical name and the DB link name (Fig.4).
Next step after sites definition is the fragmentation screen. List of accessible tables for the previously defined user is added as an auto complete on the first combobox. The second combobox suggests fragmentation types (horizontal, vertical and nested).
Derived fragmentation is transparent to user

As example, horizontal fragmentation interface provides user with the list of columns of chosen table (Fig.5). User enters fragment name and chooses hosting site and then checks columns related to this fragment (Fig.6). By default the tool keeps the last selection of columns so that the designer can affect the same fragment to multiple sites without redefining the fragment columns. If the designer needs to flush selection, a shortcut on F5 key is linked and functional.

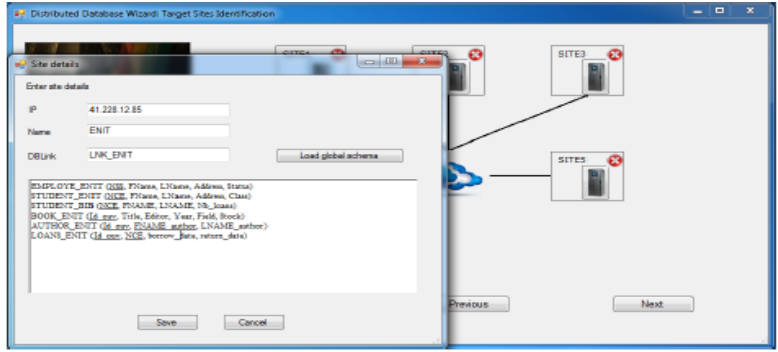

**Fig. 4.** Site definition details

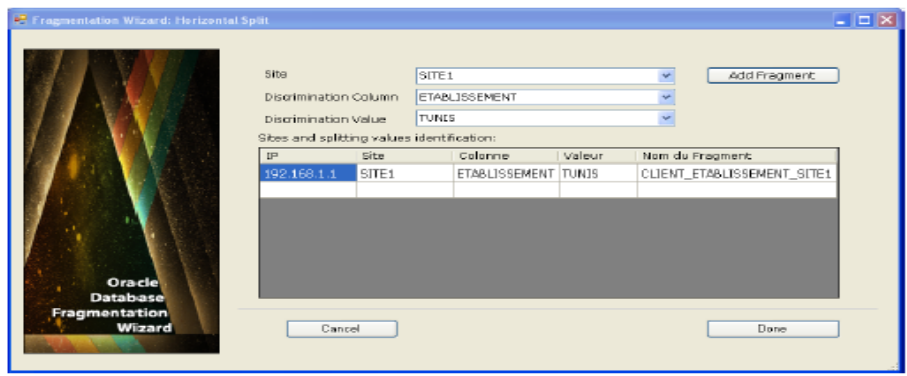

**Fig. 5.** Horizontal fragmentation sample.

Once finished the fragmentation for a table, the wizard starts an automated validation for the described configuration. Adding a fragment without a primary key is already controlled while creating the fragment (on "Add Fragment" button click). Validation screen is displayed then: The left canvas holds a fragment tree with first level nodes as sites, second level nodes as fragment names and leaves are the columns. Primary key is highlighted (orange color). In the right container, the validation report is displayed for the three validation criteria: Reconstruction, completeness then disjointness. [5]

In the end of the whole process, if the policy is validated by the wizard and designer, the tool takes in charge the transcription of visual design into SQL scripts to run on remote sites. The only necessary parameter for this operation is scripts location. Script files naming convention is as follows: [SITE_NAME]_DDB_SCRIPT.sql.

The generation process goes through all sites and generates the script to create symbolic links, then transforms into a standard fragments and commented SQL script. The field names and types are consistent with the starting table (same name and same type). Procedures, views, triggers, and the various components are then written accordingly (Figure 7).

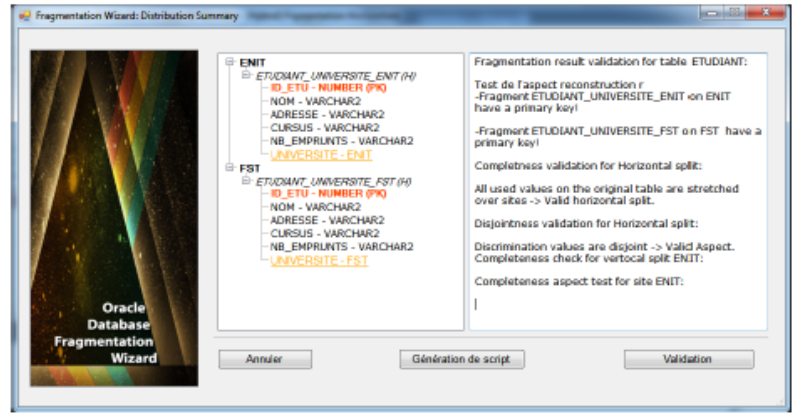

**Fig. 6.** Horizontal fragmentation validation

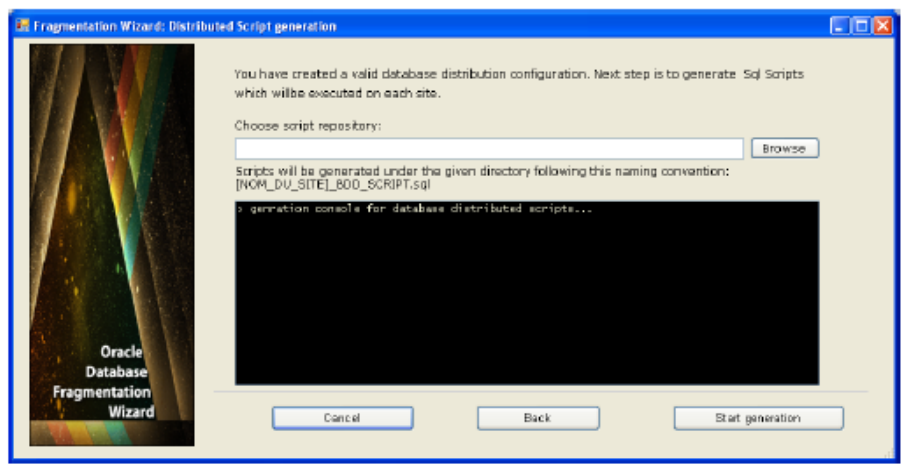

**Fig. 7.** Script generation Interface.

## 7    Conclusion

Through the study of how to implement a distributed database under Oracle and taking account of the lack of a smart Oracle assistant that allows automatic implementation of a database distribution policy, we decided to design and create an assistance layer to design and implement an Oracle DDB. The result of current work is a friendly visual wizard, which allows the translation of schemes of distributed databases designed through this tool into an executable script directly on all the nodes of the topology.

Suggested further work to improve the intelligent-DDB layer: 1) Full support of hard and software heterogeneity (Different DBMS, Different OS, Network topology) and 2) integrate performance simulator (Enable designers to anticipate bottlenecks even before implementing distribution policy, predict performance interpolation graphs based on user predefined queries).